\documentclass[showpacs,aps,prb,reprint,twocolumn]{revtex4-1}

\usepackage{commath}
\usepackage{amsmath,bm}
\usepackage{graphicx}
\usepackage{hyperref}
\usepackage{txfonts}

\draft 

\begin{document}


\title{On probable lossless surface plasma waves: a reply to criticisms}
\author{Hai-Yao Deng}
\email{haiyao.deng@gmail.com}
\affiliation{School of Physics and Astronomy, Cardiff University, 5 The Parade, Cardiff CF24 3AA, Wales, United Kingdom}
\begin{abstract}  
In a recent preprint, arXiv 2005.03716v1, G. Wegner and C. Henkel criticized my recent work on the possibility of lossless surface plasma waves. Here I refute all of their criticisms. 
\end{abstract}
\maketitle

\section{introduction}
\label{sec:1}
In the study of surface plasma waves (SPWs), I came to realize the possibility that these waves may undergo a linear instability under certain conditions~\cite{Deng2019,Deng2017a,Deng2017b,Deng2017c,Deng2020b}. I explained in Ref.~\cite{Deng2019} why this possibility had gone overlooked in the literature. This was so primarily due to a historical incorrect treatment of surface effects, which resulted in incorrect theories of SPWs. 

G. Wegner and C. Henkel (henceforth referred to as WH)~\cite{WH} disputed with my work~\cite{Deng2019,Deng2017c,Deng2017a,Deng2017b} and claimed to have found a number of errors. In what follows I show that their claim is unfounded, partly motivated by their naive mathematical manipulations while partly due to their misunderstanding of my work and some basic yet subtle physics. 

\section{On Charge Conservation}
\label{sec:2}
The first major criticism raised by WH is concerned with a relaxation term included in the equation of continuity, see Eq.~(\ref{9}) below, which they claimed was in contradiction with the conservation of charges and due to a n\"{a}ive use of Boltzmann equation. Here I show that their claim was mistaken, due to their incapability of seeing the physics behind this term. 
 
As in previous work, I here consider a metal and describe it by the jellium model. Let $-en_0$ be the charge density of the uniform positive background, where $e$ is the charge of an electron and $n_0$ is the mean electron density. The net charge density of the metal is 
\begin{equation}
\rho(\mathbf{x},t) = e \left(n(\mathbf{x},t)-n_0\right), \label{1}
\end{equation}
where $\mathbf{x}$ denotes a point in space and $t$ denotes time, and $n(\mathbf{x},t)$ gives the actual electron density. In thermodynamic equilibrium, no charges can develop in the metal without an external field,
\begin{equation}
\rho(\mathbf{x},t) = \rho_{eq} = 0.
\end{equation}
Here $\rho_{eq}$ denotes the equilibrium density. 

Suppose the metal is currently in equilibrium. We now disturb it away from the equilibrium by a \textit{transient} external field (e.g. a pulse of light), which induces a finite charge density $\rho(\mathbf{x},t)$ in it, and study the subsequent free evolution of this density. An electric current, of density $\mathbf{j}_{tot}(\mathbf{x},t)$ now flows in the metal, which in the end shall lead the metal back to equilibrium (i.e. $\rho$ to $\rho_{eq}$). The equation of continuity applies,
\begin{equation}
\partial_t \rho(\mathbf{x},t) + \partial_{\mathbf{x}}\cdot\mathbf{j}_{tot}(\mathbf{x},t) = 0. \label{3}
\end{equation}
which expresses the conservation of total charges. 

Progress is made by noting that $\mathbf{j}_{tot}$ consists of two contributions of different nature~\cite{landau}: 
\begin{equation}
\mathbf{j}_{tot} = \mathbf{j}+\mathbf{j}'. \label{4}
\end{equation}
The first contribution, $\mathbf{j}$, is driven by the electric field $\mathbf{E}$ generated by the induced charge density $\rho$. This field stems from the long-range part of the Coulomb interaction between the electrons~\cite{pines1954}. In other words,
\begin{equation}
\mathbf{j}(\mathbf{x},t) = \int d\mathbf{x}' \int dt' ~ \sigma(\mathbf{x},\mathbf{x}',t-t') \mathbf{E}(\mathbf{x}',t'), \label{5}
\end{equation}
or equivalently, in the frequency representation,
\begin{equation}\tag{5'}
\mathbf{j}_\omega(\mathbf{x}) = \int d\mathbf{x}' ~ \sigma(\mathbf{x},\mathbf{x}';\omega) \mathbf{E}_\omega(\mathbf{x}'),
\end{equation}
where $\sigma$ is the bare electrical conductivity, whose tensorial nature has been suppressed here to ease notation, and, neglecting retardation effects,
\begin{equation}
\mathbf{E}(\mathbf{x},t) = -\partial_\mathbf{x} \phi(\mathbf{x},t), \quad \partial^2_\mathbf{x} \phi + 4\pi \rho = 0
\end{equation}
with $\phi$ being the electrostatic potential. 

On the other hand, $\mathbf{j}'$ is driven by the residue short-range forces between the electrons and those due to interaction of the electrons with other scatterers, which are responsible for random electronic collisions~\cite{landau}. 

We substitute Eq.~(\ref{4}) in (\ref{3}) and obtain
\begin{equation}
\partial_t \rho(\mathbf{x},t) + \partial_{\mathbf{x}}\cdot\mathbf{j}(\mathbf{x},t) = - \partial_{\mathbf{x}}\cdot\mathbf{j}'(\mathbf{x},t) \label{7}
\end{equation}
Now, in order to assert the importance of $\mathbf{j}'$, let us proceed with a thought experiment, in which we switch off the long-range forces, that is, we set $\mathbf{j} = 0$. Under such circumstances, $\rho$ could evolve only by means of $\mathbf{j}'$, toward $\rho_{eq}$ to restore equilibrium in the end. This shows that it is essential to include $\mathbf{j}'$ for a complete description of charge dynamics in the system. 

As a natural approximation, we have in previous work~\cite{Deng2019,Deng2017c,Deng2017a,Deng2017b,Deng2020b} taken that
\begin{equation}
\partial_{\mathbf{x}}\cdot\mathbf{j}'(\mathbf{x},t) = \rho(\mathbf{x},t)/\tau, \label{8}
\end{equation}
where $\tau$ denotes a relaxation time, which is roughly the time it takes for all of the charges in a unit volume to disperse out of the volume by $\mathbf{j}'$. Using this in Eq.~(\ref{7}), we arrive at
\begin{equation}
\left(\partial_t + \frac{1}{\tau} \right)\rho(\mathbf{x},t) + \partial_{\mathbf{x}}\cdot\mathbf{j}(\mathbf{x},t) = 0. \label{9} 
\end{equation}
This equation directly follows from the approximation (\ref{8}) and is independent of the specifics of electron dynamics, be they quantum mechanical or classical. As a confirmation, one may see that it is consistent with Boltzmann equation~\cite{Deng2017c} as well as with the quantum kinetic equation~\cite{harris1972,harris1971}. 

As said in the above, Eq.~(\ref{8}) and hence Eq.~(\ref{9}) are approximate. Alternatively, one may assume a diffusion picture and invoke Fick's law 
\begin{equation}
\mathbf{j}' = -eD\partial_\mathbf{x}n = -D\partial_\mathbf{x}\rho, \label{9'}
\end{equation}
where $D \sim l^2_0/\tau$ is the electronic diffusion constant with $l_0$ being roughly the electronic mean free path. Eq.~(\ref{9'}) leads to $\partial_{\mathbf{x}}\cdot\mathbf{j}' = -D\partial^2_\mathbf{x} \rho \sim \rho/\tau$. This option describes the same physics as option (\ref{8}) and they lead to similar results.

WH failed to see the physical origin of Eq.~(\ref{9}) and wrongly criticized its use in my work~\cite{Deng2019,Deng2017c,Deng2017a,Deng2017b,Deng2020b}. They deployed a proposition by Mermin~\cite{mermin1970} to defend their reasoning, without realizing that this proposition is only a trick that has never been derived from fundamental physics. Mermin put forth his trick to ``conserve local electron number". However, why should the electron number be conserved locally, given that it is actually not conserved, i.e. the number of electrons in any volume element is constantly changing?

\section{On Current Density}
\label{sec:3}
Another major criticism raised by WH is concerned with my utilizing
\begin{equation}
\mathbf{j}(\mathbf{x},t) = \theta(z) \mathbf{J} (\mathbf{x},t), \label{10}
\end{equation}
where $\mathbf{J}$ -- whose exact physical meaning becomes clear later -- gives the current density in the bulk but not the surface of the metal, in Eq.~(\ref{9}) to establish in the metal that
\begin{equation}
\left(\partial_t + \frac{1}{\tau} \right)\rho(\mathbf{x},t) + \partial_{\mathbf{x}}\cdot\mathbf{J}(\mathbf{x},t) = -\theta'(z) J_z(\mathbf{x}_0). \label{11}
\end{equation} 
In these equations, a semi-infinite metal (SIM) occupying the half space $z\geq 0$ has been considered, $\mathbf{x}_0 = (\mathbf{r},0)$ denotes a point on the macroscopic surface with $\mathbf{r} = (x,y)$ and $\theta(z)$ is a step function. 

WH argued for the absence of the term on the right-hand-side of Eq.~(\ref{11}) by insisting that $J_z(\mathbf{x}_0) = 0$. Here I show that they have again failed to appreciate the physics, this time that behind Eq.~(\ref{10}), and have committed a common historical error related to the general issue of the so-called additional boundary conditions (ABCs). 

\begin{figure}
\begin{center}
\includegraphics[width=0.45\textwidth]{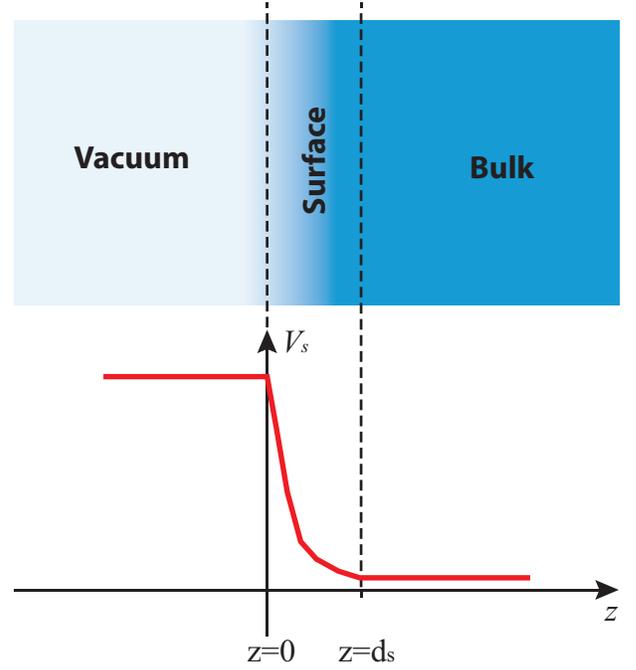}
\end{center}
\caption{Microscopic picture of a semi-infinite metal (SIM), which divides into a surface region ($0<z<d_s$) and a bulk region ($z>d_s$). In the surface region the confining potential $\partial_\mathbf{x}V_s \neq 0$, whereas in the bulk region, as in the vacuum, $\partial_\mathbf{x}V_s = 0$. \label{fig:1}}
\end{figure} 

\begin{figure*}
\begin{center}
\includegraphics[width=0.95\textwidth]{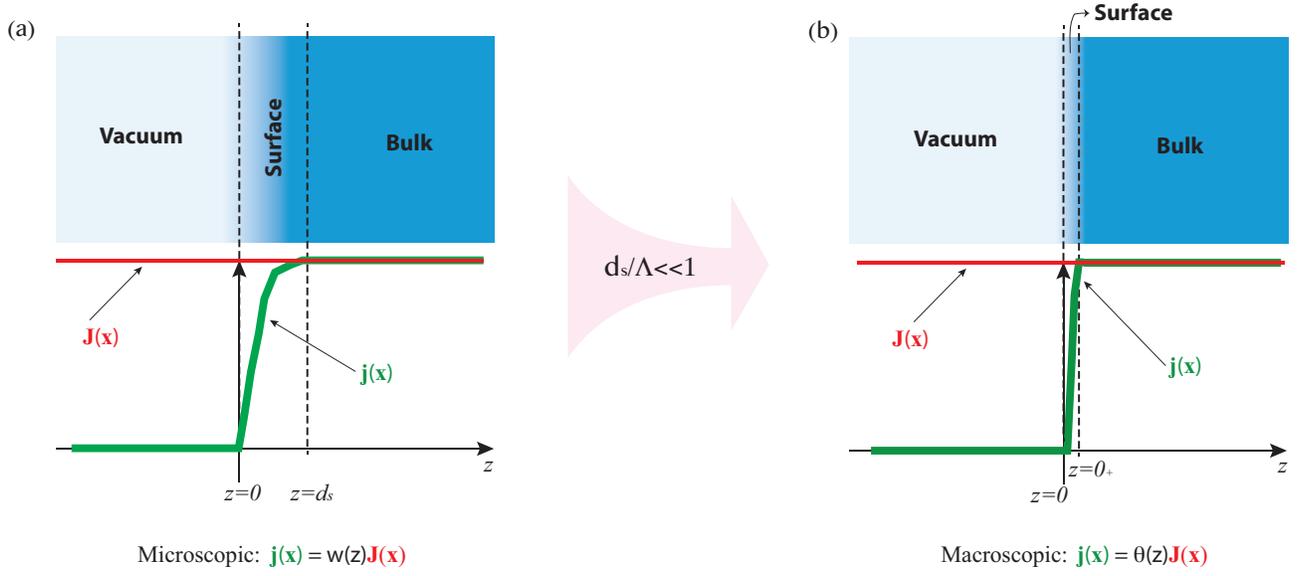}
\end{center}
\caption{Microscopic (a) and macroscopic (b) picture of the surface region. Here ``macroscopic" is meant by the same sense as the ordinary Maxwell's boundary conditions are applied, attained by conventional coarse-graining. On should note that, however thin the surface region appears, it must be there and should not be reduced to a simple geometrical plane.\label{fig:2}}
\end{figure*} 

To appreciate the physics behind Eq.~(\ref{10}), let us resort to the microscopic picture of the surface of a SIM. This picture has been briefly discussed in Ref.~\cite{Deng2019}, which WH failed to appreciate, and with more details in Ref.~\cite{Deng2020a,Deng2020b}. As sketched in Fig.~\ref{fig:1}, microscopically, the global system divides into three regions: the vacuum ($z<0$), the surface layer ($0\leq z\leq d_s$) and the bulk ($z>d_s$). Here $d_s$ is the thickness of the surface layer. The division is based on the behaviors of the confining potential $V_s(\mathbf{x})$, which prevents electrons from escaping the metal. Only in the surface region does $V_s$ vary significantly. In both the vacuum and the bulk it can be treated as constant, as sketched in Fig.~\ref{fig:1}. As electrons are prevented from moving beyond the plane $z=0$ by the confining potential (neglect quantum tunneling), the current density $\mathbf{j}$ must vanish at $z=0$, namely,
\begin{equation}
\mathbf{j}(\mathbf{x}_0,t) = 0, \label{12}
\end{equation}
where $\mathbf{x}_0 = (\mathbf{r},z=0)$ denotes a point on the plane $z=0$ and $\mathbf{r} = (x,y)$. This is the condition of impenetrability. 

In the regime of linear responses, Eq.~(\ref{5}), or equivalently, Eq.~(5'), can be used to relate $\mathbf{j}$ to the electric field $\mathbf{E}$ present in the metal by means of the bare electrical conductivity tensor $\sigma(\mathbf{x},\mathbf{x}',t-t')$, which can be calculated in many ways. For the purpose of illustration, let us for example use the Kubo-Greenwood formula to this end, 
\begin{equation}
\sigma_{\mu\nu}(\mathbf{x},\mathbf{x}';\omega) = i\hbar \sum_{m,n} \frac{f_m - f_n}{\varepsilon_n - \varepsilon_m} \frac{\hat{j}^\mu_{nm}(\mathbf{x}) \hat{j}^\nu_{mn}(\mathbf{x}')}{\hbar(\omega+i\tau^{-1}) + \varepsilon_m - \varepsilon_n}. \label{13}
\end{equation}
Here we have restated the tensorial indices, $f_m$ are the Fermi-Dirac population factors for the eigenstates $\psi_m$ of the single-particle Hamiltonian $H$, i.e.
\begin{equation}
H\psi_m = \varepsilon_m \psi_m, \quad H= H_0 + V_s,
\label{14}
\end{equation}
with $\varepsilon_m$ corresponding to the energies, and $$\hat{\mathbf{j}}_{mn}(\mathbf{x}) = \langle \psi_m|\hat{\mathbf{j}}(\mathbf{x})|\psi_n\rangle, \quad \hat{\mathbf{j}} = \left(\hat{j}^x,\hat{j}^y,\hat{j}^z\right)$$ are the matrix elements of the current density operator $\hat{\mathbf{j}}$. Needless to say, $\psi_m(\mathbf{x}_0) = 0$ so that Eq.~(\ref{12}) is fulfilled. Indeed, $\sigma(\mathbf{x},\mathbf{x}')$ vanishes for either $\mathbf{x}$ or $\mathbf{x}'$ lying in the vacuum, as it should. 

To evaluate Eq.~(\ref{13}), one must solve Eq.~(\ref{14}), which, however, involves a major obstacle, that of $V_s$ being unknown \textit{a priori}. Indeed, $V_s$ in reality can be very complicated and depends on many factors: even for samples of the same metal, $V_s$ could differ depending on how the samples are processed, stored and so on. In many existing work, one often employs a hard-wall picture for the surface, which takes $V_s$ to be zero for $z>0$ and infinity for $z\leq 0$. This amounts to taking $d_s = 0$ and totally obliterating the surface region. We shall come back to this point later on. 

Notwithstanding, we can still write down the general form of $\psi_m$ in the bulk region, where $V_s = 0$. Indeed, $\psi_m$ solves 
$$H_0\psi_m = \varepsilon_m \psi_m$$ in the bulk. Let us denote by $\Psi_m$ the eigenstates of $H_0$, i.e. $H_0\Psi_m = \varepsilon_m \Psi_m$. Then, in general we have \begin{equation}
\psi_m(\mathbf{x} \in \text{bulk}) = \Psi_m(\mathbf{x}) + \sum_{m'} t_{mm'} \Psi_{m'}(\mathbf{x}), \label{15}
\end{equation}
where $\Psi_{m'}$ and $\Psi_m$ have the same energy as $\psi_m$, and $t_{mm'}$ are amplitudes. For a crystal, in Eq.~(\ref{15}) $\Psi_m$ represents a Bloch wave impinging on the surface and the terms in the sum stand for scattered (reflected) waves. The amplitudes $t_{mm'}$ therefore describe scattering effects due to $V_s$ and can be regarded as reminiscence of surface effects in the wave function $\psi_m$ in the bulk region. In a macroscopic theory, they can be taken as phenomenological parameters. In the present work, their exact forms are irrelevant, but those interested are referred to Ref.~\cite{Deng2020c} for more details. 

The main message drawn out of the foregoing discussions is that, $\psi_m(\mathbf{x}\in\text{bulk})$ can be determined up to a set of parameters that characterize surface scattering effects. As such, it is useful to introduce a new quantity, $\Sigma(\mathbf{x},\mathbf{x}';\omega)$, defined by Eq.~(\ref{13}) but with $\hat{\mathbf{j}}_{mn}$ calculated using Eq.~(\ref{15}). In other words, $\Sigma(\mathbf{x},\mathbf{x}';\omega)$ is identical with $\sigma(\mathbf{x},\mathbf{x}';\omega)$ in the bulk region but not in the surface region. We further introduce
\begin{equation}
\mathbf{J}_\omega(\mathbf{x}) = \int_{\mathbf{x}'\in\text{metal}} d\mathbf{x}' ~ \Sigma(\mathbf{x},\mathbf{x}';\omega) \mathbf{E}_\omega(\mathbf{x}'). 
\end{equation}
It follows that
\begin{eqnarray}
&~& \mathbf{j}_\omega(\mathbf{x}\in\text{bulk}) = \int_{\mathbf{x}'\in\text{metal}} d\mathbf{x}' ~ \sigma(\mathbf{x},\mathbf{x}';\omega) \mathbf{E}_\omega(\mathbf{x}') \\ 
&~& = \int_{\mathbf{x}'\in\text{bulk}} d\mathbf{x}' ~ \Sigma(\mathbf{x},\mathbf{x}';\omega) \mathbf{E}_\omega(\mathbf{x}') \nonumber \\ &~& \quad \quad \quad \quad  \quad \quad  + \int_{\mathbf{x}'\in\text{surface}} d\mathbf{x}' ~ \sigma(\mathbf{x},\mathbf{x}';\omega) \mathbf{E}_\omega(\mathbf{x}') \\
&~& = \int_{\mathbf{x}'\in\text{metal}} d\mathbf{x}' ~ \Sigma(\mathbf{x},\mathbf{x}';\omega) \mathbf{E}_\omega(\mathbf{x}') = \mathbf{J}_\omega(\mathbf{x}),
\end{eqnarray}
 where in the last equality we have neglected
 \begin{equation}
 \int_{\mathbf{x}'\in\text{surface}} d\mathbf{x}' ~ \left(\sigma(\mathbf{x},\mathbf{x}';\omega) - \Sigma(\mathbf{x},\mathbf{x}';\omega)\right)\mathbf{E}_\omega(\mathbf{x}') \sim \frac{d_s}{\Lambda},
 \end{equation}
assuming that $\Lambda \gg d_s$, where $\Lambda$ is a macroscopic scale that is comparable to the extension of the electric field in the metal. 

To summarize, we now see that
\begin{equation}
\mathbf{j}_\omega(\mathbf{x}) 
=
\begin{cases}
\mathbf{J}_\omega(\mathbf{x}), & \text{for} ~ \mathbf{x} \in \text{bulk}, \\
0, &  \text{for} ~ \mathbf{x} \in \text{vacuum}.
\end{cases}
 \label{21}
\end{equation}
Yet, in the surface region $\mathbf{j}$ remains unknown. My strategy is to relate $\mathbf{j}$ and $\mathbf{J}$ by an extrapolation function $w(z)$, as sketched in Fig.~\ref{fig:2} (a), so that
\begin{equation}
\mathbf{j}_\omega(\mathbf{x}) = w(z) \mathbf{J}_\omega(\mathbf{x}). \label{22}
\end{equation}
The function $w(z)$ is assumed to evolve smoothly from zero in the vacuum region to unity in the bulk region,
\begin{equation}
w(z) 
=
\begin{cases}
1, & \text{for} ~ z>d_s, \\
0, &  \text{for} ~ z\leq 0.
\end{cases}
 \label{23}
\end{equation}
We have assumed that $w$ depends only on $z$, which is reasonable for a flat surface. 

At this point, one should see that, though $\mathbf{j}$ vanishes at $z=0$, as required by the condition of impenetrability (\ref{12}) and ensured by the extrapolation function $w(z)$, $\mathbf{J}$ does not in general. The reason is obvious: in the surface region the current density is given by $\mathbf{j}$ not by $\mathbf{J}$. 

Finally, we go to the macroscopic limit, see Fig~\ref{fig:2} (b). Here ``macroscopic" is meant by the same sense as the ordinary Maxwell's boundary conditions are applied. More precisely, we mean a coarse-graining process on the scale $\Lambda\gg d_s$, by which the surface region appears infinitely thin (but, of course, always there) and $w(z)$ universally degenerates with the step function $\theta(z)$. With this result, we revisit Eq.~(\ref{10}) by Eq.~(\ref{22}). As said above, $\mathbf{J}(\mathbf{x}_0)$ does not vanish in general and hence the right-hand-side of Eq.~(\ref{11}) must be present and determined self-consistently. 

In the above, to illustrate the point, I have explicitly made use of Kubo-Greenwood formula. However, this is not necessary. For example, one may illustrate the same point by a semi-classical theory based on Boltzmann equation, in which case the parameter in place of $t_{mm'}$ is the \textit{Fuchs} parameter (see Sec.~\ref{sec:5}). Actually, it has recently~\cite{Deng2020c} been shown that the electrical conductivity obtained by Kubo-Greenwood formula reduces to that based on Boltzmann equation in the semi-classical limit, and meanwhile $t_{mm'}$ reduces to the \textit{Fuchs} parameter. 

We conclude this section by mentioning two historic fallacies. WH has committed both. The first fallacy refers to the following practice: the authors of many existing work, when evaluating the electrical conductivity (and physical quantities of a similar kind), did not include $V_s$ in their dynamical equations (e.g. Schr\"{o}dinger's equation, Boltzmann's equation and Navier-Stokes equation), and hence they were computing $\mathbf{J}$ instead of $\mathbf{j}$. Nevertheless, these authors failed to recognize the subtle difference exposed in the above analysis, and mistook $\mathbf{J}$ for $\mathbf{j}$. This has led to numerous permeating errors that need to be rectified in the literature. More discussions on this are made in Secs.~\ref{sec:5} and \ref{sec:6}.  

In the second fallacy, due to the misconception leading to the first fallacy, many authors, including WH, wrongly imposed that $J_z(\mathbf{x}_0) = 0$. They failed to see that the impenetrability of the surface layer requires $\mathbf{j}$ vanish on the surface but not $\mathbf{J}$. Many authors, e.g. F. Flores and F. Garc\'{i}a-Moliner~\cite{flores1977,flores1979}, even went out of their way to fulfill this condition by contriving tricks that sound reasonable but physically ill, some of which were deployed by WH in Ref.~\cite{WH} to support their claim. It should be said that that $J_z(\mathbf{x}_0) = 0$ is actually an example of the so-called Pekar's ABC, the problem with which has been widely known~\cite{Henneberger1998}. A solution to this problem has recently been proposed by the present author~\cite{Deng2020b}.  

\section{On Energy Balance}
\label{sec:4}

\begin{figure}
\begin{center}
\includegraphics[width=0.45\textwidth]{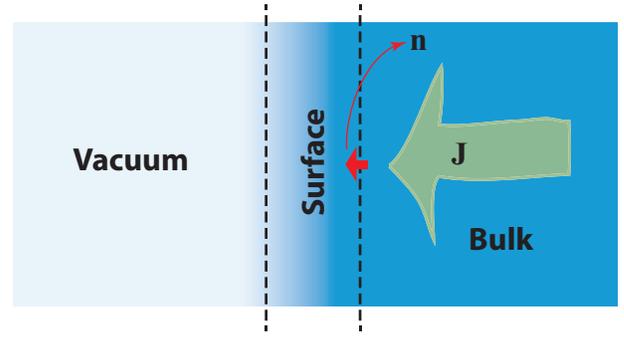}
\end{center}
\caption{A flux of charges enters the surface region leading to a change of total charges in this region at rate $\mathbf{n}\cdot\mathbf{J}$, where $\mathbf{n} = (0,0,-1)$ is the surface normal vector. This contributes to the change of the total electrostatic energy in the system. \label{fig:3}}
\end{figure} 

The third major criticism raised by WH is concerned with the energy balance equation I used in Ref.~\cite{Deng2017c}. As we shall see, in their simple-minded mathematical manipulations~\cite{WH}, WH did not bear in mind the microscopic picture and lost the surface contribution. As with their other claims, they have failed to appreciate the subtle physics for the math. Their calculations, presented in Sec. III B in their paper~\cite{WH}, were based on their incorrect energy balance equation and a modified semi-classical model due to Zaremba (see Sec.~\ref{sec:5} for detailed discussions on this point), and are hence irrelevant to my work. In the rest of this section, I reproduce the derivation of the correct energy balance equation. 

The reasoning leading to this equation is actually straightforward and briefly sketched here. Let $E_p$ be the electrostatic energy of the system. Its rate of change equals the negative of the work done on the electrons by the electric field $\mathbf{E}$, that is, 
\begin{equation}
\dot{E}_p = - \int d^3\mathbf{x} ~ \mathbf{j}(\mathbf{x},t) \cdot \mathbf{E}(\mathbf{x},t), \label{24}
\end{equation}
which can also be directly derived from the equation of continuity, as shown in Ref.~\cite{Deng2017c}. The integral involved here is split into a bulk part $P_b$ and a surface part $P_s$, with
\begin{eqnarray}
&~& P_b = \int_{\mathbf{x}\in \text{bulk}} d^3\mathbf{x} ~ \mathbf{j}(\mathbf{x},t) \cdot \mathbf{E}(\mathbf{x},t) \nonumber \\ &~& \quad \quad \quad \quad \quad \quad \quad = \int_{\mathbf{x}\in \text{bulk}} d^3\mathbf{x} ~ \mathbf{J}(\mathbf{x},t) \cdot \mathbf{E}(\mathbf{x},t),
\end{eqnarray}
and
\begin{equation}
P_s = \int_{\mathbf{x}\in \text{surface}} d^3\mathbf{x} ~ \mathbf{j}(\mathbf{x}) \cdot \mathbf{E}(\mathbf{x}).
\end{equation}
Note that $-P_s$ can also be interpreted as the rate of change of the electrostatic energy in the surface region, which in the macroscopic limit is given by
\begin{equation}
-P_s = \int d^2\mathbf{r} ~ \phi(\mathbf{x}_0,t) \mathbf{n}\cdot\mathbf{J}(\mathbf{x}_0,t) = - \int d^2\mathbf{r} ~ \phi(\mathbf{x}_0,t) J_z (\mathbf{x}_0,t). \label{27} 
\end{equation}
Here $\mathbf{n} = (0,0,-1)$ is the unit normal vector of the surface. Physically, $\mathbf{n}\cdot\mathbf{J}(\mathbf{x}_0,t)$ gives the number of total charges that enter the surface region per unit area per unit time, see Fig.~\ref{fig:3}. Combining Eqs.~(\ref{24}) - (\ref{27}), we find
\begin{equation}
\dot{E}_p = -\int_{\mathbf{x}\in \text{bulk}} d^3\mathbf{x} ~ \mathbf{J}(\mathbf{x},t) \cdot \mathbf{E}(\mathbf{x},t) - \int d^2\mathbf{r} ~ \phi(\mathbf{x}_0,t) J_z (\mathbf{x}_0,t),\label{eb}
\end{equation}
which is the energy balance equation derived in Ref.~\cite{Deng2017c}. It has been rigorously shown that~\cite{Deng2017b}, the SPW amplification rate obtained using this energy balance equation is identical with that obtained by the equation of motion. This further confirms the self-consistency of Eq.~(\ref{eb}). WH failed to appreciate this contribution that ultimately derives from the surface charges.

\section{On The Semi-Classical Model}
\label{sec:5}
The criticism raised by WH on my use of the semi-classical model (SCM) was motivated by their insisting $J_z(\mathbf{x}_0,t) \equiv 0$, which, as shown in Sec.~\ref{sec:3}, is incorrect. All of their calculations were based on a modified semi-classical model by Zaremba~\cite{zaremba1974}, which was invented to satisfy the incorrect condition $J_z(\mathbf{x}_0,t) \equiv 0$ and will be discussed later, and are therefore irrelevant to my work. As such, this criticism does not deserve further analysis. Nevertheless, I would like to take this opportunity to rectify a few misconceptions about the standard SCM. WH, like many others, was unable to avoid any of these misconceptions. 

For the sake of completeness, here I briefly recapitulate the basics of the model. The starting point of SCM is the linearized Boltzmann equation under the relaxation approximation. For a SIM, we may assume without loss of generality for any field quantity $\mathcal{F}$ the time and planar coordinates dependence $\sim e^{i(kx-\omega t)}$, i.e. we write
\begin{equation}
\mathcal{F}(\mathbf{x},t) = \mathcal{F}(z) \sim e^{i(kx-\omega t)}. 
\end{equation}
Hereafter, this prescription is implicit with any field quantity that displays explicitly only $z$-dependence. With it, Boltzmann's equation may be written as 
\begin{equation}
\left(\partial_z + \frac{1}{\lambda}\right)g(\mathbf{v},z) + ef'_0(\varepsilon) \frac{\mathbf{v}\cdot\mathbf{E}(z)}{v_z} = 0. \label{29}
\end{equation}
Here $f_0(\varepsilon)$ is the Fermi-Dirac distribution function, where $\varepsilon = m\mathbf{v}^2/2$ is the kinetic energy of an electron with velocity $\mathbf{v} = (v_x,v_y,v_z)$ and mass $m$, $f'_0(\varepsilon) = \partial_\varepsilon f_0$, $\lambda = iv_z/(\bar{\omega} - kv_x)$ with $\bar{\omega} = \omega+i\tau^{-1}$, and $g(\mathbf{v},z)$ denotes the non-equilibrium distribution of the electrons. Equation~(\ref{29}) can be solved and a unique solution is obtained when supplemented by two boundary conditions set at $z\rightarrow \infty$ and $z=0$, i.e.
\begin{equation}
g(\mathbf{v},z\rightarrow \infty) = 0, \quad g(\mathbf{v},z=0) = p g(\mathbf{v}_-,z=0), \label{30}
\end{equation} 
where in the second condition $\mathbf{v} = (v_x,v_y,v_z>0)$, $\mathbf{v}_- = (v_x,v_y,-v_z)$, and $p$ is called the specularity parameter that was first introduced by K. Fuchs~\cite{fuchs1937}. Physically, $p$ is usually considered as the fraction of electrons incident upon the surface to get specularly reflected back. Obviously, it characterizes surface effects and plays the role of $t_{mm'}$ in Kubo-Greenwood formula~\cite{Deng2020b}. 

Now we are ready to discuss the misconceptions. The first is related to the fact that, many researchers, without thinking, habitually use the solution $g(\mathbf{v},z)$ to compute the charge and current densities. Namely, they take 
\begin{equation}
\tilde{\rho} = \left(\frac{m}{2\pi\hbar}\right)^3 e \int d^3\mathbf{v} ~ g(\mathbf{v},z) \label{31}
\end{equation}
as the charge density and 
\begin{equation}
\tilde{\mathbf{J}} = \left(\frac{m}{2\pi\hbar}\right)^3 e \int d^3\mathbf{v} ~ \mathbf{v} g(\mathbf{v},z) \label{32}
\end{equation}
as the current density. Unfortunately, this is incorrect. The reason is simple: Eq.~(\ref{29}) does not contain the confining potential $V_s$, and hence, as explained in Sec.~\ref{sec:3}, $g(\mathbf{v},z)$ is invalid in the surface region. Consequently, both $\tilde{\rho}$ and $\tilde{\mathbf{J}}$ are invalid in the surface region. Obviously, $\tilde{\mathbf{J}}$ is identical with $\mathbf{J}$ but not $\mathbf{j}$ [c.f.~Sec.~\ref{sec:3}]. 

This brings us to a widespread incorrect practice in the study of SPWs. Many~\cite{harris1972,harris1971,flores1977} employed the laws of electrostatics to express $g$ as a functional of $\tilde{\rho}$ and substitute it in Eq.~(\ref{31}) to obtain an equation involving only $\tilde{\rho}$, which is then taken as the basic equation for SPWs. For example, WH followed exactly this incorrect procedure in carrying out analysis in Sec. II C in their paper~\cite{WH}. 

Another misconception is connected to the boundary condition at $z=0$, i.e. the second relation in Eq.~(\ref{30}), which assumes that a fraction $p$ of incoming electrons are specularly reflected back by the surface. As a result, the current density calculated by Eq.~(\ref{32}), which -- as aforementioned -- gives $\mathbf{J}$, does not vanish at the surface. This fact, in light of the analysis presented in Sec.~\ref{sec:3}, is natural and self-evident. However, innumerable researchers have been puzzled by it, due to their inability to penetrate the physical situation illuminated in Sec.~\ref{sec:3}, and many of them, e.g. Zaremba~\cite{zaremba1974}, invented physically unfounded tricks, some of which were deployed by WH, to get rid of this feature so that $J_z$ was artificially made to vanish at the surface. 

One should see that Eq.~(\ref{30}) constitute the most natural and general conditions that are allowed by Eq.~(\ref{29}), which is an first-order linear ordinary differential equation. It was first pointed out by J. Bardeen~\cite{bardeen1958} and has recently been explicitly shown that they emerge directly from a quantum mechanical theory~\cite{Deng2020b}. 

Back to Zaremba, whose point of view was adopted by WH, even if we tolerate for a while his incorrect imposition about $J_z$, there is still an obvious absurdity with his model, which assumes that incoming currents be totally balanced by outgoing currents so that no charges accumulate on the surface region. To see this, let us consider a beam of electrons normally incident on the surface at velocity $v_{in}$. For the sake of definiteness, suppose that this beam consists of 100 electrons. For simplicity, let us assume that all of these electrons are diffusely reflected, and that these diffusely reflected electrons move almost parallel to the surface so that their normal velocity is very small, say $v_{diff}$. As such, to satisfy Zaremba's model, one must have $100 v_{in}/v_{diff}>100$ diffusely reflected electrons. This, of course is absurd. At this point, it should be clear that Zaremba's model cannot conserve the total number of electrons. WH blindly adopted this model without seeing this fallacy. 

\section{On The Hydrodynamic Model and related issues}
\label{sec:6}
In Ref.~\cite{Deng2019}, I reasoned that the often-quoted SPWs, which are based on the condition that $J_z(\mathbf{x}_0) = 0$, within the hydrodynamic model (HDM) are incompatible with those derived from the local Drude model (LDM), which was already been noticed long ago~\cite{hdm}. WH dismissed my conclusion, claiming that HDM made a counter example of my proof that SPWs cannot exist in macroscopic models obeying $J_z=0$. 

My arguments for the incompatibility are multifold. In the first place, I wish to point out that, as already emphasized in Ref.~\cite{Deng2019}, the incompatibility between HDM subjected to $J_z(\mathbf{x}_0) = 0$ and LDM is self-evident, because in the latter the additional condition that $J_z(\mathbf{x}_0) = 0$ is certainly violated. Actually, LDM presumes that $\mathbf{J}_\omega(\mathbf{x}) = \sigma_{LDM}(\omega) \mathbf{E}_\omega(\mathbf{x})$, which implies that $J_{\omega,z}(\mathbf{x}_0) = \sigma_{LDM}(\omega)E_{\omega,z}(\mathbf{x}_0) \neq 0$. Here $\sigma_{LDM}$ is Drude conductivity. 
As such, however one manipulates his solutions, this incompatibility cannot be removed. WH failed to appreciate this point. 

One should also be reminded that HDM calculates $\mathbf{J}$ not $\mathbf{j}$. This is clear from the starting equation of HDM,
\begin{equation}
n_0m\left(\partial_t + \frac{1}{\tau}\right) \mathbf{v}(\mathbf{x},t) = n_0 e \mathbf{E}(\mathbf{x},t) - mv^2_0 \partial_\mathbf{x}n(\mathbf{x},t), \label{33}
\end{equation}
Here $\mathbf{v}$ is the velocity field of the electron fluid and $v_0$ is a parameter that goes to zero in the LDM limit. This equation does not contain any confining potential $V_s$ and is thus valid only in the bulk of the metal. Hence, $en_0\mathbf{v}$ gives $\mathbf{J}$ not $\mathbf{j}$ and again, as discussed in Sec.~\ref{sec:3}, there is no reason to impose $J_z(\mathbf{x}_0) = 0$. WH had failed to see this point as well. 

In addition, WH dismissed my proof -- given in Ref.~\cite{Deng2019} -- that the widely-quoted SPW solution by HDM is a different type of surface wave. My proof goes like this: following the standard procedures, I inserted the ansatz $\rho(z) = e(n-n_0) = \rho_0 e^{-\kappa z}$ in Eq.~(\ref{33}) and then, using the law of electrostatics and the equation of continuity, subjected to the condition $v_z = 0$ at $z=0$, found that $\kappa$ diverges as the parameter $v_0$ goes to zero (the LDM limit). This implies that, the total amount of charges in the solution vanishes in the LDM limit, that is, $Q = \int^\infty_0 dz~ \rho(z) = \rho_0/\kappa \rightarrow 0$ for $v_0 \rightarrow 0$. WH disputed this point, essentially contending that $Q$ should be kept a constant when taking this limit. However, even if $Q$ is kept a constant, the LDM is not recovered in the limit, because $J_z(0)$ vanishes due to the boundary condition adopted. 

In closing this section, some remarks are made as to why SPWs should not exist in macroscopic models with $J_z(\mathbf{x}_0) = 0$. Firstly, one should see that this is obvious with the LDM. The continuity equation for this model yields 
\begin{equation}
\left(4\pi \sigma_{LDM} - i\bar{\omega}\right) \rho = - \theta'(z) J_z(0). \label{35}
\end{equation}
This would give $\rho = 0$ if $J_z(0) = 0$,  unless $4\pi \sigma_{LDM} - i\bar{\omega} = 0$ corresponding the excitation of volume plasma waves. No SPWs would exist unless $J_z(0)$ is allowed to be finite. Indeed, it is exactly this term that gives rise to SPWs in LDM. Secondly, as is easy to see generally and has been demonstrated explicitly~\cite{beck1971,gerhardts1983}, if $J_z$ vanished at the surface, no charges would flow in and out of the surface layer (see Fig.~\ref{fig:3}). As a result, no surface charges could build up, let alone the existence of SPWs. This is actually also inferrable from WH's own analysis, Eq.~(20) in their paper~\cite{WH}. 

Finally, it is worth bringing about the following qualitative argument. If $J_z(\mathbf{x}_0) = 0$, then $\rho(z)$ would vanish both at $z=0$ and $z\rightarrow \infty$. This would mean that $\rho(z)$ should exhibit a peak somewhere. Macroscopically, there is no length scales other than the SPW wavelength to locate the peak. At long wavelengths, the only reasonable possibility is to have $\rho(z)$ vanishing almost everywhere; otherwise, the charges would spread throughout the entire metal. Exceptions to this argument might occur where there is a microscopic length scale that can be used to locate the peak, such as in microscopic theories; in such case, the microscopic length scale effectively demarcates a surface layer, and, as expected, $\mathbf{J}$ should be understood as the current outside this layer. This argument will be elaborated elsewhere. 

\section{Other objections raised by WH}
\label{sec:7}
In this section, we briefly address other criticisms raised by WH. Most of these have already lost their relevance in light of the analysis given in preceding sections, but for the sake of completeness I briefly mention them. Throughout this section, page numbers refer to the paper~\cite{WH} by WH. 

\subsection{On the sign of $J_z(0)$ in LDM}
On page 3, just above the last paragraph of the left column, WH claimed to spot an error in one of my papers~\cite{Deng2017c}. In that paper, I used $J_z(0) = -2\pi \rho_s$, where $\rho_s$ is the areal density of surface charges in a SPW. WH claimed it should be $2\pi \rho_s$, which is $J_z(0_+)$. The discrepancy obviously comes from their mistaking $J_z(0_+)$ for $J_z(0)$. Again WH lost the physics (i.e. the surface layer, however thin, has two sides located at $z=0$ and $z=0_+$) for the math. 

\subsection{On instability of Fermi sea.} 
On page 4, just above Sec.~I (C), WH wrote ``\textit{One may raise the question why in that case the Fermi sea of filled electronic levels should become unstable, since it is constructed as the state of the lowest energy for a fixed charge density}". They contend that Fermi sea be the state of lowest energy and could not be destabilized. This is, of course, incorrect. Fermi sea would be the ground state of a metal only if electron-electron interactions were neglected. Otherwise, it might well get unstable, superconductivity and magnetism being notable examples! In our case, it is obviously the Coulomb interactions that drive a surface charge density wave instability.  

\subsection{On the approximate form of $\mathcal{H}$}
I showed~\cite{Deng2019,Deng2017c,Deng2017a,Deng2017b} that the charge density in SIM satisfies the following equation
\begin{equation}
\int^\infty_0 dq'~\left(\mathcal{H}(q,q') - \bar{\omega}^2\delta(q-q')\right) \rho_{q'} = S, \quad S = i\bar{\omega} J_z(0), \label{37}
\end{equation}
where $\rho_q$ is the cosine Fourier transform of $\rho(z)$, that is,
\begin{equation}
\rho(z) = \int^\infty_0 dq ~\rho_q\cos(qz). \label{rho}
\end{equation}
As $\mathcal{H}$ determines the properties of volume plasma waves, which should not be strongly affected by the presence of boundary, I take as an approximation that 
\begin{equation}
\mathcal{H}(q,q') \approx \Omega^2 \delta(q-q'), \label{38}
\end{equation}
where $\bar{\omega}^2 - \Omega^2 = 0$ gives the dispersion of volume plasma waves. WH criticized me   of dropping $\rho'(0)$ in obtaining Eq.~(\ref{38}), where $\rho'(z) = \partial_z\rho(z)$. This criticism sounds strange. Firstly, as I said here and in every of my paper on the subject, Eq.~(\ref{38}) is an approximation, and the way to go beyond this has also been discussed in Ref.~\cite{Deng2017b}. If they were not content with this approximation, WH should solve Eq.~(\ref{37}) in its exact form. Secondly, the effects of the dropped terms have already been thoroughly discussed in Refs.~\cite{Deng2017c,Deng2017a,Deng2017b}, where it was shown that such effects are higher-order in $kv_F/\omega_p$. Finally, it is quite reasonable to take $\rho'(0) = 0$ for any realistic profile of $\rho(z)$, which  can never contain $\rho_q$ with $q>q_c$, where $q_c$ is of the order of an inverse of a lattice constant. As such, $\rho(z)$ must smoothly evolve away from $z=0$, implying that $\rho'(0) = 0$ by Eq.~(\ref{rho}). WH cited $e^{-\kappa z}$ to counteract this reasoning, but they did not realize that this function does not apply at $z\sim0$ in any realistic situation with a cut-off in $q$. 

\subsection{On the separation of surface and bulk charges}
In writing their Eq.~(20) on page 5 in their comment~\cite{WH}, WH was evidently motivated to consider charges in the surface region and those in the bulk region separately, the former with areal density $\rho_s$ and the latter with density $\rho_b$. This is in principle fine. The problem, however, lies with their parsing the current density $\mathbf{j}$ into a bulk part and a surface part, which are written, in their notation, $\mathbf{J}_b$ and $\mathbf{J}_s$, respectively, where, according to WH, $\mathbf{J}_s$ must be localized within the surface region. Unfortunately, as is evident from the analysis given in Sec.~\ref{sec:3}, in general there is no way to split $\mathbf{j}$ into two disconnected parts, with one being localized in the surface. Actually, it does not make any sense to do that: current varies smoothly throughout the system. The easiest way to demonstrate this is to use LDM, by which one has $\partial_\mathbf{x}\cdot \mathbf{J} = 4\pi \sigma_{LDM} \rho$, which is totally localized within the surface, though $\mathbf{J} = \sigma_{LDM} \mathbf{E}$ can by no means be split as WH imagined. 

I must here make it clear that, the $\mathbf{J}_b$ and $\mathbf{J}_s$ introduced by WH have nothing to do with the same symbols introduced in my work~\cite{Deng2019,Deng2017c,Deng2017a,Deng2017b,Deng2020b}. In my work, $\mathbf{J}_b$ and $\mathbf{J}_s$ are defined as follows
\begin{equation}
\text{Deng:} \quad \mathbf{J}_{b,s} = \int d^3\mathbf{x}' \sigma_{b,s}(\mathbf{x},\mathbf{x}') \mathbf{E}(\mathbf{x}'),
\end{equation}
where $\sigma_b$ and $\sigma_s$ are the conductivity for the infinite system and the deviation from it. In other words, I write $\Sigma = \sigma_b + \sigma_s$, where $\sigma_b$ is calculated for an infinite system and $\sigma_s$ is the remaining. Note that $\sigma_b$ is uniquely defined and so is $\sigma_s$. I have \textbf{NEVER} stated, and there is \textbf{NO} reason to believe, that $\mathbf{J}_s$ is localized in the surface.  

The foregoing analysis also categorically undermines the self-consistency of the two-type model invented by Bedeaux and Vlieger~\cite{bv2004} and adopted by others~\cite{hh2012} including WH (Sec.~IV in Ref.~\cite{WH}). I do not see any point to further analyze WH's analysis based on this model. 

\subsection{On symmetry breaking effects}
Lastly, I would like to clarify what I mean by ``symmetry breaking effects", by which WH was obviously upset, probably because they did not take into account the context within which this phrase was used in my papers. To be specific, let us look at Table 1 in Ref.~\cite{Deng2019}. There the phrase obviously refers to whether $G_s$ exists or not. In all the models except for SCM, $G_s$ vanishes, which is why I stated that SCM displays symmetry breaking effects but other models do not. Note that, I was not saying that there is no symmetry breaking effects in those models; rather, what I said was that such effects, if any, were not manifest in $G_s$. 

\subsection{On SPW loss rate}
Back to SCM, in previous work~\cite{Deng2019,Deng2017a,Deng2017b} I showed that physical causality requires $$\gamma_0 = \Im(\bar{\omega})>0,$$ to which WH apparently agreed (see Sec. II B in their paper~\cite{WH}). It follows that, by definition, the SPW loss rate is given by
\begin{equation}
\gamma = - \Im(\omega) = 1/\tau - \gamma_0.
\end{equation}
Now if $1/\tau$ drops below $\gamma_0$, SPWs could amplify. To this result, WH disagrees for unknown reasons. This result can in principle be studied experimentally. 

\subsection{On numerical convergence}
Vaguely WH also questioned the numerical convergence in regard to a cut-off $q_c$ used in integrals over $q$. However, I have checked meticulously the convergence achieved upon increasing $q_c$, see for example Ref.~\cite{Deng2020b}. Actually, such convergence is warranted: the integral over $q$ is generally of this type, $$\int^\infty_0 \frac{dq ~ k}{k^2+q^2} ... = \int^\infty_0 \frac{d \tilde{q}}{1+\tilde{q}^2} ...,$$ which automatically suppresses the contribution from $\tilde{q} = q/k \gg 1$. This effectively introduces a cut-off.  

\section{Conclusions}
\label{sec:8}
In conclusion, I have shown that WH failed to appreciate some basic yet subtle physics regarding metal surfaces. Their criticisms are hence irrelevant. I contend that SPWs could be made lossless, in principle, as shown in my previous work. 

We should see that with SPWs there are many misconceptions, many having been taken for granted by the community. Unfortunately, WH could not avoid any of them. 

As to be demonstrated elsewhere, my approach to SPWs has applications in many physical problems other than SPWs. For example, it resolves the long-standing issue of ABCs. In my approach, boundary conditions, including Maxwell's ones, can be completely bypassed, just as in microscopic theories. Together with E. Muljarov, I have recently solved the celebrated exciton polariton problem, which intrigued Pekar to introduce his famous but physically incorrect ABCs. Our results, to be published elsewhere, perfectly agree with the exact microscopic theory, which lends strong support to the basic theory I developed in Refs.~\cite{Deng2019,Deng2017a,Deng2017b,Deng2020a,Deng2020b}.

\end{document}